\documentclass{jltp}
\usepackage{graphics}
\usepackage{bm}

\title{Cosmic Problems for Condensed Matter Experiment}

\author{Tanmay Vachaspati}

\address{
CERCA, Physics Department,\\
Case Western Reserve University,\\
Cleveland, OH 44120, USA.
}

\runninghead{T.~Vachaspati}{Cosmic Problems for Condensed Matter Experiment}

\begin{document}

\maketitle

\begin{abstract}
Condensed matter analogs of the cosmological environment have raised 
the hope that laboratory experiments can be done to test theoretical 
ideas in cosmology. I will describe Unruh's sonic analog of a black 
hole (``dumbhole'') that can be used to test Hawking radiation, and 
some recent proposals on how one might be able to create a dumbhole 
in the lab. In this context, I also discuss an experiment already done 
on the Helium-3 AB system by the Lancaster group.
\end{abstract}

Cosmology is a unique science where observations of the
current or recent universe are used to infer about the
very early universe. Furthermore, the physical ideas used
to extrapolate back to the early universe are based on
ideas that were developed in a non-cosmological setting.
Surprisingly most of these ideas work extremely well. Yet
there are also ideas that have no experimental verification
but are used on the basis of theoretical extrapolation.
``Cosmology in the lab'' is an emerging research area
to search for analogs of the cosmological environment
to enable experimental tests of ideas in cosmology.

Cosmological problems are of two types. The first type contains
problems that are based on physics in which the cosmological
environment is not crucial. It is just that the particular
problem is more relevant to cosmology than to some terrestrial 
system. The outcomes of phase transitions fall into this class 
of problems. In cosmology, remnants of a phase transition, 
such as topological defects, are crucial since they could be 
observed today and lead to important clues about an earlier epoch. 

Problems of the second type are strongly based on the gravitational
environment. For example, these could involve quantum effects in
the presence of a ``horizon''. In models of cosmic inflation, quantum 
field theoretic effects on scales larger than the cosmic horizon 
are supposed to have led to density fluctuations that then grew 
into galaxies. However, these quantum field theoretic effects are 
built upon the success quantum field theory (QFT) has 
had on very small (atomic) scales. Systems with horizons have not
been constructed in the lab so far, and there are no experimental
tests of QFT on superhorizon scales. Another system with a horizon 
is a black hole. Particles within the black hole cannot classically
escape to the region outside the horizon. However, quantum effects 
are supposed to modify this picture, leading to Hawking radiation 
from the black hole~\cite{Haw74}. It would be very desirable to find 
a way to experimentally test these ideas.

Research on ``cosmology in the lab'' should be understood strictly 
as an attempt to simulate the 
cosmological environment in laboratory systems. It should not 
become an attempt to equate cosmological (or gravitational) 
phenomenon to some feature of a particular laboratory system. 
For example, cosmology in the lab cannot say if there is an 
underlying atomistic structure to spacetime. However, if there 
was an atomistic theory of spacetime, condensed matter experiments 
might prove to be useful analogs. From my perspective, cosmology 
in the lab is similar to numerical studies of systems carried out 
on a computer. If the simulation is accurate, experiments can tell 
us things about cosmology that we would otherwise have no hope of 
finding out.

\begin{figure}
{\centerline{{\scalebox{0.50}{\includegraphics{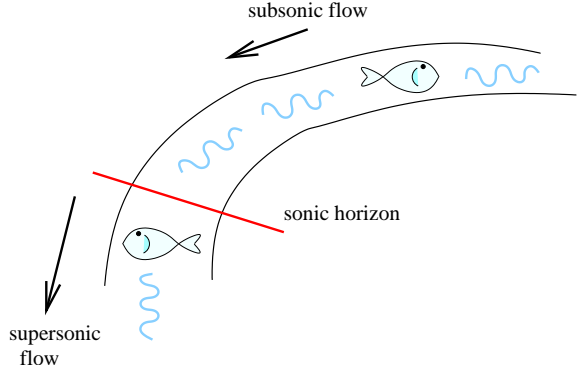}}}}}
\caption{\small \label{fig1} Unruh's vision of a waterfall as a sonic 
black hole or ``dumbhole''. The fish in the subsonic flow region cannot 
hear the screams of the fish in the supersonic region because 
the emitted sound travels too slowly to propagate upstream. The 
fish upstream sees a sonic horizon at the location where the fluid 
velocity becomes supersonic.}
\end{figure}

In 1981 Unruh~\cite{Unr81} proposed an experimental analog of 
a black hole. He considered a fluid that is flowing at subsonic speed
in the upstream region and at supersonic speed in the downstream region 
(see Fig.~\ref{fig1}). Any fish in the downstream region cannot 
send sound signals to the fish in the upstream region. Therefore 
there is a sonic event horizon. Classically, the fish upstream 
will receive no sound from the sonic hole {\it i.e.} the hole
is ``dumb''. With quantum effects taken into account, however, 
there will be Hawking radiation of sound from the horizon of
the dumbhole just as there is Hawking radiation of light from 
a black hole.

\

To put Unruh's dumbhole on a more quantitative footing, the fluid 
is assumed to obey the irrotational, inviscid, Navier-Stokes 
equations: 
\begin{eqnarray}
{\bm \nabla} \times {\bm v} = 0 \ , \ \ 
\frac{\partial \rho}{\partial t} + 
{\bm \nabla}\cdot (\rho {\bm v}) = 0 \nonumber \\
\frac{\partial {\bm v}}{\partial t} + {\bm v}\cdot {\bm \nabla} {\bm v}
 = -\frac{1}{\rho}{\bm \nabla}p - {\bm \nabla}\Phi 
\end{eqnarray}
where $\Phi$ denotes an external potential ({\it e.g.} gravitational
potential). The pressure ($p$) is assumed to be a function of the
density ($\rho$) alone. Fluctuations around some background flow will 
correspond to sound. If $\rho_0$ and ${\bm v}_0$ denote the background
flow, 
$$
\rho = \rho_0 + \delta \rho
$$
$$
{\bm v} = {\bm v}_0+{\bm \nabla}\phi
$$
Then it can be shown that
\begin{equation}
\nabla_\mu \nabla^\mu \phi =
\frac{1}{\sqrt{-g}} 
\partial_\mu ( \sqrt{-g} g^{\mu \nu} \partial_\nu \phi ) = 0
\label{waveeq}
\end{equation}
where the metric, $g_{\mu\nu}$, experienced by the sound is given by:
\begin{equation}
ds^2 = 
  ( c_s^2 - v_0^2 ) dt^2 + 2 v_0 dt dr - dr^2 - r^2 d\Omega^2 
\label{metric}
\end{equation}
This is the Painlev\'e-Gullstrand-Lema\^itre form of a black hole 
metric if we assume a spherically symmetric, stationary, convergent, 
background flow. The horizon is at $v_0 = c_s$, that is, at
the location where the fluid velocity equals the sound speed. 

The next step is to include quantum field theoretic effects. For
this we have to go back to Eq.~(\ref{waveeq}), find the mode
functions, and compare their behavior near the horizon to that
in the asymptotic region. This is the usual Hawking calculation
and we will not show it here. We simply
state Unruh's result for the Hawking temperature~\cite{Unr81}:
\begin{equation}
T_H = \frac{1}{2\pi} \frac{\partial v_0}{\partial r} \biggr | _{hor}
    = (3\times 10^{-7} \ K) 
      \left [ \frac{c_s}{300 {\rm m/s}} \right ]
      \left [ \frac{1 {\rm mm}}{R} \right ]
\label{unruhestimate}
\end{equation}
where $R$ is the distance over which $c_s$ changes.

Unruh's estimate highlights the difficulty of the problem. We
need to accelerate the fluid by $300$ m/s over a distance of
$1$ mm to get a temperature of a mere $\sim 10^{-7}$ K. In 
units of the acceleration due to gravity, the required fluid 
acceleration is $ \sim $ $10^7 g$! Achieving such an enormous
acceleration is impractical given that we also want the fluid
to remain very cold ($\mu$K), so that we are able to 
detect the Hawking radiation. There are all sorts of instabilities 
that set in under such extreme conditions for all known 
systems~\cite{Unr02}.

What is the alternative? A ray of hope emerges when one
considers Visser's generalization~\cite{Vis98} of Unruh's result.
Unruh in his pioneering contribution only considered a fluid in
which the sound speed is fixed. Visser generalized the
setting to include the possibility that the sound speed may
vary within the fluid. Then he found:
\begin{equation}
T_{sH} = \left ( \frac{\hbar}{2\pi k_B} \right )
                 \frac{\partial}{\partial r}(c_s-v) ~ \biggr |_{hor}
\label{visserTH}
\end{equation}
where $c_s = c_s (t, {\bm x})$ is the sound speed and the derivative
is evaluated at the location of the horizon ($v=c_s$).

The ray of hope~\cite{Vac03} is that we need not have huge 
gradients in $v$. Instead we might be able to arrange $c_s$ to vary 
very rapidly.  An extreme case would be where $v\approx 0$ everywhere 
but the properties of the fluid change at a boundary that is moving. 
The fluid is stationary with respect to the container but is moving 
with respect to the boundary. 

This was also the underlying idea in a proposal made by Jacobson and 
Volovik~\cite{JacVol98}
where they considered a moving texture or domain wall in $^3$He. 
The sound speed within the wall is different from that outside. And
the wall needs to be moved relatively slowly with respect to the
container for there to be Hawking radiation. One complication there
is that the order parameter varies within the wall in such a way
that there is also radiation due to other effects.

\begin{figure}
\centerline{\scalebox{0.50}{\includegraphics{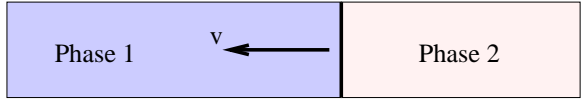}}}
\caption{\small \label{fig2} A supercooled or superheated system with 
two phases can be a black hole analog if the phase boundary propagates 
with subsonic speed with respect to Phase 1 but supersonic speed with
respect to Phase 2 ($c_1 > v > c_2$). In the rest frame of the boundary, 
the fluid is moving to the right at a speed $v$ that is subsonic to
the left of the boundary and supersonic to its right.
}
\end{figure}

The basic experimental set-up seems very simple (see Fig.~\ref{fig2}).
There is a tube with the fluid in one phase on one side and a
second phase on the other side. We assume that the sound speed
in phase 2 is less than the sound speed in phase 1: $c_1 > c_2$. 
If the phase boundary propagates from phase 2 into phase 1 at a
speed $v$ such that $c_1 > v > c_2$, sound from within the phase 2
region cannot enter the phase 1 region. Alternately, in the rest frame 
of the boundary, the fluid is flowing with velocity $v$ toward the phase 2
side, and the Unruh set-up is exactly duplicated. The phase boundary
is the location of the sonic horizon. Since the fluid is not
moving with respect to the container, there won't be any instabilities
due to surface effects. Although, the phase boundary does move with 
respect to the container and one needs to worry if this will lead to 
instabilities.

\

The Hawking temperature for the propagating phase boundary can 
be calculated in the rest frame of the boundary. Then the metric 
is of the Painlev\'e-Gullstrand-Lema\^itre
form (Eq.~(\ref{metric})) and one can directly use the 
calculations in Refs.~\onlinecite{ParWil00,Volbook} in which 
Hawking radiation is viewed as the tunneling of ingoing
particles from inside the black hole to outgoing particles
on the outside. The advantage of this approach is that it is 
largely system independent. It does require, however, that the
particles be massless both inside and outside the black
hole. Let us now outline this calculation following 
Ref.~\onlinecite{Volbook}.

In both the exterior and interior regions of the sonic hole 
(Phases 1 and 2 in Fig.~\ref{fig2}), the dispersion relation 
is assumed to be that for a massless fermionic particle. In the
fluid rest frame:
\begin{equation}
E(p) = \pm c_1 p \ , \ \ {\rm exterior}
\end{equation}
\begin{equation}
E(p) = \pm c_2 p \ , \ \ {\rm interior}
\end{equation}
For convenience we will write these equations as:
\begin{equation}
E(p) = \pm c p
\end{equation}
where $c(x)$ is $c_1$ in the exterior region and $c_2$ in
the interior region. The transition in $c(x)$ is assumed to
be sharp compared to any other length scale of interest
but $c(x)$ is still assumed to be continuous.

As for any fermion, the dispersion relation has states with both 
negative and positive energy. The negative energy levels are all
occupied and form the Dirac sea.

\begin{figure}
\centerline{\scalebox{0.50}{\includegraphics{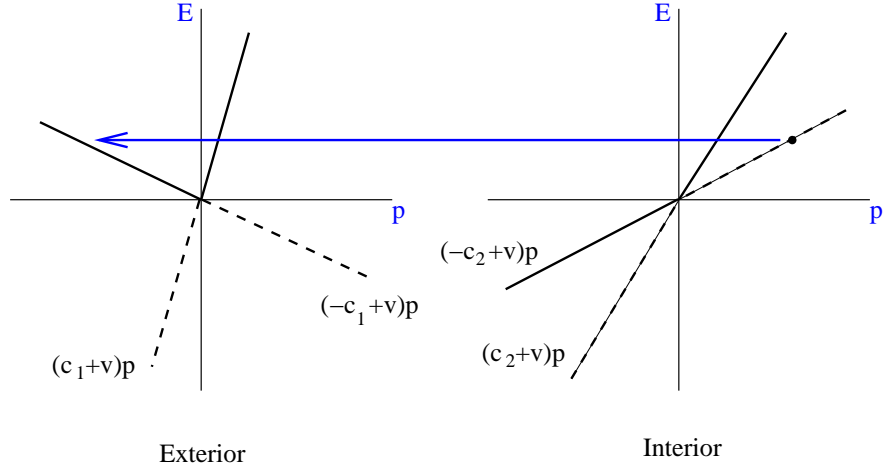}}}
\caption{\small \label{fig3} The dispersion relation outside (left diagram) 
and inside (right diagram) the dumbhole. Due to the supersonic flow inside
the dumbhole the occupied levels in the Dirac sea (dashed lines) emerge 
and get positive energy. Now particles in the emerged states can tunnel 
to an unoccupied state in the exterior region~\cite{Volbook}.}
\end{figure}

If we work in the rest frame of the phase boundary, the fluid
moves to the right with speed $v$ in Fig.~\ref{fig2}. Now the
velocity of a quasiparticle, defined as ${\bm \nabla}_p E$,
is shifted by the fluid velocity. So the dispersion relation 
becomes:
\begin{equation}
E(p) = (\pm c +v) p 
\end{equation}
The usual situation is when $v < c_2 < c_1$. Then the filled levels (Dirac
sea) of both branches of the dispersion relation have negative energy. 
What happens if $v > c_2$? Then one branch of the Dirac sea in the 
interior region has positive energy as illustrated in Fig.~\ref{fig3}. 
This is what makes it possible for particles of energy $E$ in the 
interior vacuum to emerge out of the dumbhole. As shown in Fig.~\ref{fig3},
it is the particles that have velocity directed to the right (further
into the interior of the dumbhole), that can tunnel out onto the branch
for the left-moving particles in the exterior region and then escape
out to the asymptotic region.

The energy is related to the momentum of the particle via 
$E = (\pm c +v) p$. We are interested in the particle that 
is outgoing in the exterior region, therefore we choose the $-$ sign.

The Hawking temperature is calculated by finding the tunneling
rate, which can be obtained from the imaginary part of the action:
\begin{equation}
{\rm Im} (S) = {\rm Im} \int_{-\infty}^{+\infty} p(x) dx = 
        {\rm Im} \int_{-\infty}^{+\infty} \frac{E}{v-c(x)} dx
\end{equation}
The function $v-c(x)$ has a zero at the location of the phase
boundary ($x=0$). So we need to go over to the complex plane, deform 
the contour of integration around the pole at the origin, and then
evaluate the integral using the residue theorem. This gives:
\begin{equation}
{\rm Im}(S) = \frac{\pi E}{|c'|} 
\end{equation}
where $|c'|$ is the magnitude of the derivative of the sound speed 
at the location of the horizon. The rate at which particles can 
tunnel to the exterior region is given by 
${\rm exp}(2~ {\rm Im}S) = {\rm exp}(E/T_H)$ where 
$T_H = |c'|/2\pi$ is the Hawking temperature.

Then the formula for the Hawking temperature is the same as 
in Eq.~(\ref{unruhestimate}). The Hawking temperature is estimated 
to be:
\begin{equation}
T_{sH} = 0.04 \ {\rm K} 
      \biggl ( \frac{\delta c}{300 {\rm {m/s}}} \biggr )
      \biggl ( \frac{100 {\mbox{\AA}}}{\xi} \biggr )
\label{boundaryTsH}
\end{equation}
where, $\xi$ is the thickness of the phase boundary.

Note that $v$ enters the estimate since it determines the 
location of the horizon at which gradient of $c$ is
evaluated. However, we will assume that $T_{sH}$ is (roughly) 
independent of $v$ as long as it satisfies $c_1 > v > c_2$.

The hardest problem, of course, is to find an experimental
realization of a dumbhole. Since the Hawking temperature is 
very low, the system must either be a fluid or a solid. Only 
a few fluids are known at this temperature. Solids may be
used too if there is a suitable melting transition. Another 
possibility is to use Bose-Einstein condensates that have been
discussed in other related contexts in 
Refs.~\onlinecite{GarAngCirZol00,GarAngCirZol01,FedFis03}.

\begin{figure}
\centerline{\scalebox{0.50}{\includegraphics{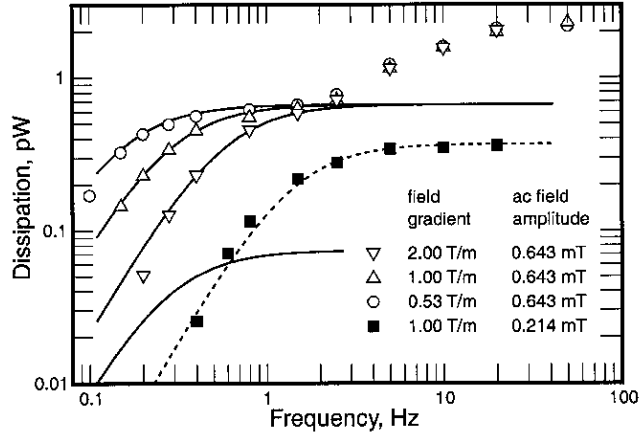}}}
\caption{\small \label{fig4} The data showing the dissipation from the 
moving AB interface as a function of the driving frequency as
given in Ref.~\onlinecite{Baretal00}. See text for further explanation.}
\end{figure}

One particularly intriguing system is superfluid $^3$He since 
the AB phase boundary has been studied quite extensively both
experimentally and theoretically. Indeed the AB interface has 
been made to oscillate by applying suitable magnetic fields, and 
the Lancaster group has also measured the radiation from the 
oscillating phase boundary~\cite{Baretal00}. The experimental setup 
is essentially that shown in Fig.~\ref{fig2}, except that the container 
is vertical.  The fluid is kept at 150 $\mu$K and 0 bar pressure.
At such low temperatures there are few excitations present and the
system is essentially in its vacuum state.  A non-uniform magnetic 
field is applied along the vertical such that it is stronger in 
the lower part of the container than in the upper part. 
This causes the lower part to be in the A-phase and the upper part to
be in the B-phase. The AB phase boundary is at the location where
the magnetic field attains a critical value, $B_{AB}$. Next, a small 
oscillating magnetic field is applied along the vertical so that the 
total magnetic field is:
\begin{equation}
B(t,z) = B_0 (z) + B_{AC} \sin (\omega t)
\end{equation}
The location of the AB phase boundary in the absence of dissipation 
can be be found by setting $B(t,z) = B_{AB}$. Due to the oscillating 
component of the magnetic field, the equilibrium location of the 
interface is time-dependent: $z_0 = a \sin (\omega t)$. The amplitude 
of oscillation is given by: 
$a = B_{AC}/\nabla B_0$. (The gradient of $B_0$ is roughly constant 
in the region where the phase boundary oscillates.) The frequency of 
oscillation of the phase boundary can be changed by changing the 
frequency, $\nu =\omega/2\pi$, of the applied oscillating field. 
The quasiparticle radiation is detected by a vibrating wire resonator 
placed at the upper end of the container. The data for the radiated 
power versus oscillation frequency is shown in Fig.~\ref{fig4}.

The behavior of the experimental curves in Fig.~\ref{fig4} 
are explained phenomenologically as follows. Assuming a linear
restoring force and dissipative motion, the equation of motion
for $z(t)$, the position of the interface, is:
\begin{equation}
-k (z- z_0 (t)) - \gamma \frac{dz}{dt} = 0
\label{eqofmotion}
\end{equation}
where $z_0 (t) = a \sin (\omega t)$ ($a = B_{AC}/\nabla B$) 
is the equilibrium position of the interface. Then,
\begin{equation}
v(t) = \frac{dz}{dt} = v_0 \cos (\omega t - \phi ) + O(e^{-\kappa t})
\label{velocity}
\end{equation}
where
\begin{equation}
\kappa = \frac{k}{\gamma} \ , \ \ 
v_0 = \frac{a \kappa \omega}{\sqrt{\kappa^2 + \omega^2}} \ , \ \ 
\tan \phi = \frac{\omega}{\kappa}
\end{equation}

Therefore $v_0 \propto \omega$ for $\omega << \kappa$
and $v_0$ is independent of $\omega$ for $\omega >> \kappa$.
The average power dissipated over an oscillation time
period is 
\begin{equation}
P = \gamma \langle v^2 \rangle
\end{equation}
and this grows quadratically with $\omega$ for $\omega < \kappa$ 
and is constant for $\omega > \kappa$. This is consistent with
the log-log plot of the data for $\nu < 1$ Hz in 
Fig.~\ref{fig4} for the larger value of $B_{AC}$, and over the whole 
range of $\nu$ for the smaller value of $B_{AC}$ provided we treat 
$\gamma$ as a fitting parameter that depends on $B_{AC}$. 

There are three unexplained features of the data as discussed in
Ref.~\onlinecite{Baretal00}.
\begin{enumerate}
\item Theoretical estimates of $\gamma$ are small by several orders of
magnitude \cite{LegYip90}.
\item For the smaller value of $B_{AC}$, the required value of $\gamma$
is about 5 times smaller than what is needed to fit the data for larger
$B_{AC}$. If $\gamma$ is calculated from the $B_{AC}=0.643$ mT data,
and used to predict the $B_{AC}=0.214$ mT curve, one gets the lower
solid curve shown in Fig.~\ref{fig4} and this does not even come close
to matching the data.
\item The data clearly shows an unexpected increase in dissipation
at frequencies $\nu > 1$ Hz when $B_{AC} = 0.643$ mT.
\end{enumerate} 

Can Hawking radiation play any role in the increased dissipation
seen for $\nu > 1$ Hz when $B_{AC} = 0.643$ mT? The data does seem 
to indicate a new source 
of dissipation at high frequencies. The velocity of the interface
at $\nu \sim 1$ Hz is about $1$ cm/s. This also happens to be close
to the quasiparticle velocity in the A phase in the direction orthogonal
to the interface ($c_A = 3$ cm/s for certain excitations as explained
below). It is also less than the quasiparticle velocity in the B phase 
($c_B = 55$ m/s). Furthermore, the power emitted in Hawking radiation 
using the temperature in Eq.~(\ref{boundaryTsH}) is of order $1$ pW 
and is in the range seen. Hawking radiation is only expected during
the part of the oscillation when the velocity of the interface lies
in the suitable range: $c_B > v > c_A$. As this duration increases,
we expect the amount of Hawking radiation to grow proportionally.

At $\nu \sim 1$ Hz in the experiment, the simple model for the motion
of the interface must break down. This could happen in two ways. The 
restoring force might get non-linear corrections, or the dissipation
parameter $\gamma$ could get some velocity dependent corrections.
Purely for illustrative purposes, let us assume that the
form of the velocity (Eq.~(\ref{velocity})) continues to hold even 
at large $\omega$ but with a different amplitude:
\begin{equation}
v = \alpha \omega \cos (\omega t - \tilde \phi )
\end{equation}
The amplitude $\alpha$ is assumed to be an $\omega$ independent 
parameter in the frequency range of interest. The exact form of $v$ 
is not crucial for us. For example, the power of $\omega$ in the 
amplitude could be different from 1.
                                                                                
Let us now assume that Hawking radiation at temperature $T_{sH}$
(Eq.~(\ref{boundaryTsH})) is emitted when $v ={\dot z} > c_A$. Then, 
for $j$ ``light'' species of radiation, this gives:
\begin{equation}
P_H = j \sigma_s T_{sH}^4 A \left ( \frac{\delta t}{\tau} \right )
  = j \sigma_s T_{sH}^4 A \frac{1}{\pi}
  \cos^{-1}
 \left ( \frac{c_A}{2\pi \alpha \nu} \right )
\end{equation}
where $\delta t/\tau$ is the fraction of time for which $v > c_A$.
Note that the formula holds only for radiation for which
the mass is less than the Hawking temperature. Since
$T_{sH} \approx 3 \rm{mK} > \Delta_B \approx 1.7 \rm{mK}$
this is self-consistent. (If the Hawking temperature is less than
the mass of the particles, the radiation is exponentially suppressed.)
Let us write
\begin{equation}
P_H = \frac{P_0}{\pi} \cos^{-1} \left ( \frac{\nu_*}{\nu} \right )
\end{equation}
where
\begin{equation}
P_0 \approx 116
        \left ( \frac{j}{2} \right )
         \left ( \frac{d}{4.3 {\rm{mm}}} \right )^2
         \left ( \frac{\delta c_s}{60 {\rm{m/s}}} \right )^2
         \left ( \frac{100 {\mbox{\AA}}}{\xi} \right )^4
         {\rm {pW}}
\end{equation}
and $d$ is the cell diameter ($4.3$ mm in the experiment).
The value of $P_0$ is in the observed range if $\xi \sim 300$ \AA.
                                                                                
Can we estimate the critical frequency $\nu_*$? In our
illustrative model 
\begin{equation}
\nu_* = \frac{c_A}{2\pi \alpha}
\end{equation}
If $\alpha =B_{AC}/\nabla B_0|_{\rm{interface}}$ then the
values of $\nu_*$ are shown in the following Table.

\begin{center}
\begin{tabular}{|c|c|c|}
\hline
$\nabla B_0$ T/m & $B_{AC}$ mT & $\nu_*$ Hz \\
\hline
2.00&0.643&14.8    \\
1.00&0.643&7.4     \\
0.53&0.643&3.9     \\
1.00&0.214&22.2    \\
\hline
\end{tabular}
\end{center}
Clearly the values of $\nu_*$, the critical value of the frequency
at which Hawking radiation starts, do not agree with data. The 
experiment shows anomalous radiation starting at some critical
{\it frequency}, whereas Hawking radiation would start at some critical
{\it velocity}. On the other hand, there is no reason to adopt 
$\alpha = B_{AC}/\nabla B_0$ except that this holds at low frequencies.
The Hawking radiation explanation can only work if $\alpha$ is 
independent of $\nabla B_0$ at high frequencies. However, as we 
will now see, the interpretation in terms of Hawking radiation
has additional theoretical issues.

The main difficulty with the interpretation is that, in our earlier 
estimate, we have assumed that the sound quanta are massless on either 
side of the interface, whereas this is not true in the AB system.
The quasiparticles on the A-phase side of the interface indeed
have a linear dispersion relation very close to the nodal point.
This can be seen from the full dispersion relation in the rest
frame of the fluid~\cite{Volbook1}:
\begin{equation}
E({\bm p}) = \pm \left [ v_F^2 (|{\bm p}| - p_F)^2 +
        \frac{\Delta_A^2}{p_F^2} ({\hat {\bm l}}\times {\bm p})^2
        \right ]^{1/2}
\end{equation}
where $v_F \approx 55$ m/s is the Fermi velocity and $p_F =m^* v_F$ 
is the Fermi momentum with $m^* \approx 3m_{He-3}$, and 
$\Delta_A \approx 2.02 k_B T_c$ where $T_c$ is the transition
temperature from normal to superfluid phase~\cite{VolWol90}. 
At zero pressure and strong magnetic field $T_c \sim 1$ mK.
It is easy to check that $E({\bm p})$ vanishes if 
${\bm p} = \pm p_F {\hat {\bm l}}$. Hence the dispersion
relation has a node. Now consider excitations near the node, 
${\bm p} = \pm p_F {\hat {\bm l}} + \delta {\bm p}$, with
\begin{equation}
\frac{|\delta {\bm p}|}{p_F} \ll \frac{\Delta_A}{p_F v_F} \sim 10^{-3} 
\end{equation}
and with momenta perpendicular to ${\hat {\bm l}}$: 
$\delta {\bm p} \cdot {\hat {\bm l}} =0$~\footnote
{Note that the ${\hat {\bm l}}$ vectors on the AB interface mostly
lie in the plane of the interface. Hence excitations that cross the
interface necessarily have non-vanishing momenta perpendicular
to ${\hat {\bm l}}$.}.
For these excitations the dispersion relation is linear,
\begin{equation}
E  = \pm c_A |\delta {\bm p}|
\label{Adispersionlow}
\end{equation}
with $c_A = \Delta_A/p_F \approx 3$ cm/s. Therefore we can expect
a dumbhole to form whenever the interface velocity exceeds $3$ cm/s.

On the B-phase side, the dispersion relation in the rest frame of
the fluid is:
\begin{equation}
E({\bm p}) = \pm \biggl [ \{ \epsilon ({\bm p}) - \mu \} ^2 +
                      \Delta_B^2 \frac{{\bm p}^2}{p_F^2}
                        \biggr ]^{1/2}
\end{equation}
where $\epsilon ({\bm p}) = {\bm p}^2/2m^*$, 
$\mu = \epsilon ({\bm p}_F)$, and the B-phase gap 
$\Delta_B \approx 1.76 k_B T_c$ 
with $T_c \approx 0.93$ mK at zero pressure and magnetic field.
Since $E$ only depends on ${\bm p}^2$ and it does not vanish at 
${\bm p}^2 = p_F^2$, there is no node in the dispersion relation. 
In fact, for $p= |{\bm p}| \approx p_F$ we have:
\begin{equation}
E({\bm p}) = \pm \biggl [ v_F^2 (p - p_F )^2 + \Delta_B^2 \biggr ]^{1/2}
\label{Bdispersionlow}
\end{equation}
where we also assume 
\begin{equation}
\biggl ( \frac{\Delta_B}{\mu} \biggr ) ^2 \ll
\biggl ( \frac{p-p_F}{p_F} \biggr ) \ll 1
\end{equation}
Note that $(\Delta_B/\mu )^2 \sim 10^{-6}$. 

\begin{figure}
\centerline{\scalebox{0.50}{\includegraphics{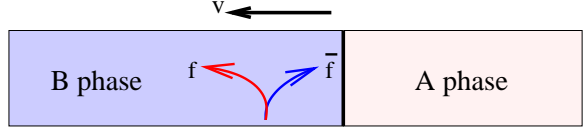}}}
\caption{\small \label{fig5}
Vacuum fluctuations in the B phase can produce an $f{\bar f}$ pair,
say with $f$ having positive energy and ${\bar f}$ having negative 
energy. ${\bar f}$ falls into the dumbhole, and if its velocity is 
very low, cannot escape back out into the B-phase region. $f$ escapes 
and forms Hawking radiation.}
\end{figure}

Since the B-phase dispersion relation is not of the usual relativistic 
form $E = \pm \sqrt{{\bm p}^2 c^2 + m^2 c^4}$, it is not possible to 
think of the quasiparticles as propagating on a metric with just the 
usual second derivative kinetic term. However the black hole analogy 
need not break down! There is still a sonic horizon from within which 
quasiparticles moving at 3 cm/s in the A-phase region cannot escape. 
This is an important ingredient for a dumbhole (see Fig.~\ref{fig5}). 
The present situation is more complicated though, because not all
quasiparticles in the A-phase move at 3 cm/s. Only the quasiparticles
close to the nodal point have this velocity. So a negative energy
particle that falls through the interface and into the A-phase region
cannot escape back to the B-phase region only if its velocity in the
A-phase region is of order 3 cm/s.

The dispersion relations given above are in the fluid rest frame. In 
the interface rest frame, the fluid is moving to the right with velocity 
${\bm v}$. Quasiparticle velocities in the interface rest frame will be 
shifted by ${\bm v}$ with respect to the velocities in the fluid rest 
frame. Since velocity is defined by ${\bm \nabla}_p E$, the dispersion
relation in the interface rest frame is obtained by adding 
${\bm v} \cdot {\bm p}$ to the dispersion relation in the fluid rest 
frame. As before, part of the Dirac sea on the A-phase side of the 
interface will emerge into the positive energy region. Furthermore,
the energy of some occupied states in the A-phase side is the same
as that of unoccupied states in the B-phase side. So the basic set-up
of the linear dispersion relation calculation given above is available.
The technical difficulty is that the equation $\epsilon = E(p,c(x))$ 
cannot as simply be inverted to get $p(E,c(x))$. Hence let us construct 
a toy model for the dispersion relation that captures some of the
essential features of the full model. 

Consider the one-dimensional dispersion relation:
\begin{equation}
E(p) = \pm\sqrt{c^2 (p-P)^2 + \Delta^2} ~ + ~ v p
\label{toydispersion}
\end{equation}
where $c=v_F$, $\Delta =\Delta_B$ in the B-phase region
and $c=c_A$, $\Delta =0$ in the A-phase. We will take 
$P=p_F \equiv P_B$ in the B-phase and $P=0$ in the A-phase, but
also keep open the possibility that $P_B=0$. The functions $c(x)$, 
$\Delta (x)$ and $P(x)$ only vary appreciably within the interface. 
This dispersion relation goes over to Eq.~(\ref{Adispersionlow}) in 
the A-phase (with $v=0$) and likewise Eq.~(\ref{Bdispersionlow}) 
in the B-phase\footnote{Though the linear dispersion in the A-phase
is due to the $\Delta_A^2$ term and not due to the kinetic term
as in the toy model.}. Note that the dispersion
relation is written in the rest frame of the interface in which
the fluid is moving to the right with speed $v$. Also, we are
looking at the one dimensional problem where the momentum is
only along the $x$ direction.

Inverting Eq.~(\ref{toydispersion}) we get,
\begin{equation}
q \equiv p - P = \frac{\epsilon v \pm 
             \sqrt{\epsilon^2 v^2 - (v^2-c^2)(\epsilon^2-\Delta^2)}}
                {(v^2-c^2)}
\label{inverted}
\end{equation}
where $\epsilon \equiv E-vP$. The discriminant should be positive 
in the asymptotic regions for the particle to be physical (and
not ``under the barrier''). This imposes a constraint on the 
velocity:  
\begin{equation}
v^2 > c^2 \left ( 1 - \frac{\epsilon^2}{\Delta^2} \right )
\label{discriminant}
\end{equation}
In the A-phase region, $\Delta =0$, and the condition is trivially
satisfied as seen from Eq.~(\ref{inverted}). In the B-phase, the 
condition reads:
\begin{equation}
v^2 > v_F^2 \left [ 1 - \frac{(E-v P_B)^2}{\Delta_B^2} \right ]
\label{constraint}
\end{equation}
Hawking radiation can only occur if:
\begin{equation}
E > \Delta_B \sqrt{1-\frac{v^2}{v_F^2}} ~ + ~ v P_B
\label{first}
\end{equation}
or else if:
\begin{equation}
E < - \Delta_B \sqrt{1-\frac{v^2}{v_F^2}} ~ + ~ v P_B
\label{second}
\end{equation}
We assume that the condition in Eq.~(\ref{discriminant}) is
satisfied everywhere, including within the interface. If this
is not the case, the discussion will be more involved.

In the experiment, $v \sim c_A \sim 10^{-3}v_F$ and so the
square root factor is essentially 1. Therefore either
$E > \Delta_B + v P_B$ or else $E < -\Delta_B + v P_B$. Note
that we are interested in $v > \Delta_A/p_F = c_A$ and
since $\Delta_A \approx \Delta_B$, we also have
$v p_F > \Delta_B$. So the thresholds for $E$ are both positive
if $P_B =p_F$ (but not if, for example, $P_B =0$). 
The first possibility is one that is expected since the minimum
energy of quasiparticles in the B-phase is $\Delta_B$ in the
fluid rest frame. The second possibility corresponds to the
emergence of the Dirac sea in the B-phase when viewed in the
rest frame of the interface. Let us discuss both possibilities
in some more detail.

At the large values of the energy required by the first possibility
(Eq.~(\ref{first})), 
our toy model starts becoming suspect and a more realistic calculation 
is called for. However, if we proceed with the toy model under the 
assumption that it still gives the correct qualitative behavior, 
we can calculate the Hawking temperature. The calculation of the 
action follows the linear case described earlier. There is a pole 
at $v=c(x)$ and this gives the only contribution to the imaginary 
part of the action. However, the process requires a branch change.
The particles on the A-phase side are on the branch with the
$-$ sign in Eq.~(\ref{toydispersion}) while, after tunneling,
the particles would end up on the branch on the B-phase side
that has the $+$ sign. Further analysis is required to determine
if the branch change can take place.

\begin{figure}
\centerline{\scalebox{0.50}{\includegraphics{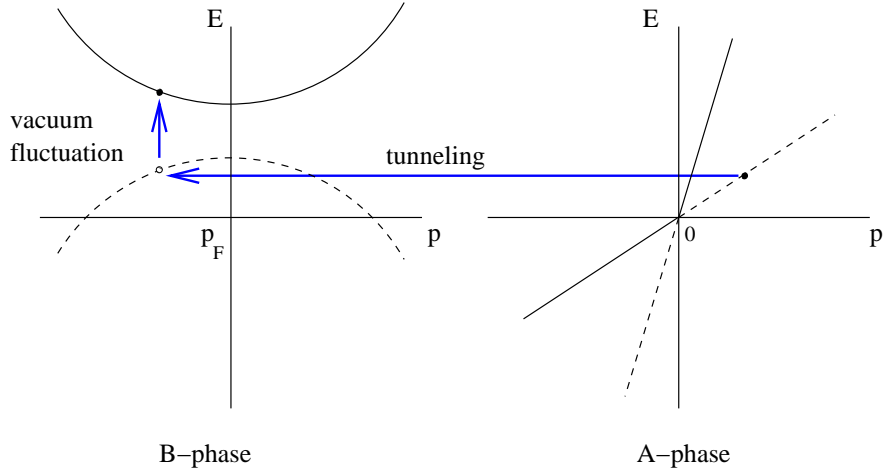}}}
\caption{\small \label{fig6}
The figure shows the schematics of the dispersion relations in 
the toy model for the A- and B- phases. Dashed curves denote the
filled states of the Dirac sea. If there is a vacuum 
fluctuation in the B-phase, a hole can be created in the lower
branch, which can then be re-filled by a particle tunneling out 
of the A-phase. The particle that had jumped onto the upper branch 
during the vacuum fluctuation then escapes to infinity as Hawking
radiation. This process is identical to that shown in Fig.~\ref{fig5}.
}
\end{figure}

Now we discuss the second possibility (Eq.~(\ref{second})). In 
this case, the particle
from the A-phase may tunnel to a state in the Dirac sea of the
B-phase. However, the tunneling cannot take place since the corresponding 
state in the B-phase Dirac sea is occupied. Yet the process may still 
be important. Suppose there is a vacuum fluctuation in the B-phase in 
which a particle from the Dirac sea gets excited to the upper branch. 
Normally the particle would fall back into the hole in the Dirac sea 
within a time allowed by the uncertainty principle. However, in the 
present situation, a particle from the A-phase can tunnel into the 
B-phase and fill up the hole before the original particle has had a 
chance to fall back. Then there is no hole left for the particle in the 
upper branch to return to and it must escape as a real particle (see 
Fig.~\ref{fig6}). This literally corresponds to the process shown 
in Fig.~\ref{fig5} where vacuum fluctuations in the B-phase create a pair 
of positive and negative energy fermions and the negative energy particle 
falls into the dumbhole. So the second possibility given by Eq.~(\ref{second}) 
might indeed be relevant. 

The calculation of the Hawking temperature 
proceeds as before. The tunneling rate is calculated by finding 
the imaginary part of the action obtained by integrating $q(x)$. 
Note that the $+$ sign in Eq.~(\ref{inverted}) must be chosen for
there to be a pole in $q(x)$. The tunneling rate must be multiplied
by the probability of having a hole of energy $E$ on the B-phase side.
As long as this factor is not an exponential, the Boltzmann factor 
will be given by the tunneling rate alone. Since the structure
of the pole is the same as in the linear calculation given above,
the Hawking temperature will still be $T_{sH} = |c'|/2\pi$ in the
interface rest frame. Note that since the tunneling occurs between
the branches with a $-$ sign in Eq.~(\ref{toydispersion}), $E$ must
lie in the range (for $P_B = p_F$):
\begin{equation}
0 < E < v p_F - \Delta_B  
\end{equation}
where we are assuming $v \ll v_F$. Therefore, in the fluid rest frame, 
the radiated particle can only have energy ($E_r$) in the interval:
\begin{equation}
\Delta_B < E_r < v p_F 
\end{equation}
This restriction on the range of energy of radiated particles will
change if branch changing processes can occur. Then one can imagine
several other processes as well. These should be investigated.

The above discussion is based on a toy model of the AB system and
suffers from the danger that it is inaccurate in some essential way. 
In particular, since the last term in Eq.~(\ref{toydispersion}) comes
from ${\bm v}\cdot {\bm p}$, and the component of ${\bm p}$ along
${\bm v}$ is taken to be small on the A-phase side, perhaps it
would have been better to replace the term by $v (p-P(x))$. 
Then, with $P_B = p_F$ and for small $p-p_F$, the B-phase Dirac sea 
would not emerge into the positive energy region and the only 
possibility for tunneling would be via branch changing processes.

Clearly it is important to do a more thorough calculation and 
with a realistic model of the AB interface. Such an effort would 
be aided by the early calculations on the dissipation from the AB 
interface~\cite{LegYip90}, the profile of the order parameter within 
the interface~\cite{SchWax89}, and the scattering of quasiparticles 
off the interface~\cite{SchWax92}. To connect with experiment, one
would also need a handle on the interface motion at high frequencies and 
on other effects that could potentially explain the anomalous radiation. 

Experimentally, from the perspective of studying Hawking radiation, it 
would be desirable to obtain the spectrum of the emitted radiation, and 
to be able to correlate the emitted power with the dynamics (say, 
the velocity) of the interface.

If it turns out that Hawking radiation from the AB interface of $^3$He
is highly suppressed (or absent), we will be left with the challenge of 
finding a quantum black hole analog. This is an exciting quest. It is
probably also our only hope for experimentally testing current ideas 
on quantum black holes and other cosmological problems.

\section*{ACKNOWLEDGMENTS}

I would like to thank the organizers of the conference on 
Quantum Phenomena at Low Temperatures (ULTI III Users Meeting, 2004) 
in Lammi, Finland, for giving me the opportunity to participate. 
I am grateful to Arnie Dahm, Shaun Fisher, Harsh Mathur, 
George Pickett, Nils Schopohl and Grisha Volovik for very informative 
discussions. This work was supported by the DOE (US).

\end{document}